\documentstyle{article}

\begin{document}
\begin{center}
{\Large\bf Missing Doublet Multiplet as the Origin of the \\
        Doublet-Triplet Splitting in SUSY SU(6)}

\vspace{3mm}

{\bf I.\ Gogoladze  A.\ Kobakhidze  and  Z.\ Tavartkiladze
}
\vspace{3mm}

{\it Institute of Physics of Georgian Academy of Sciences,
Tamarashvili 6, Tbilisi 380077, \\ Republic of Georgia
}
\end{center}

\begin{abstract}
In a $SU(6)$ gauge theory we found the irreducible representation
(175-plet) which does not contain the Higgs doublet.
Using this representation we construct two SUSY $SU(6)$ models in which
the doublet-triplet splitting occurs naturally, without fine tuning.
The crucial role is played
by the ``custodial" global $SU(2)_H$ in combination with
discrete or continuous $R$ symmetries.
\end{abstract}
\vspace{2mm}

\section{Introduction}

The problem of gauge hierarchy is concomitant to any Grand
 Unified Theories (GUT).
While solving this problem two questions must be understood:
1.~Why the large scale of GUT ($M_{G}\sim 10^{16}~$ GeV)
is not felt by the electroweak scale $m_{W}\sim 100$ GeV?
In other words, what is the reason that $SU(2)_{W}\times U(1)_{Y}$ symmetry
breaking scale is so small and stable against radiative corrections?
2.~How does the Higgs doublets remain light when their color
triplet partners from the same irreducible representation (IRREP) must
be superheavy in order to avoid fast proton decay?
The latter problem is known as the
doublet-triplet (DT) splitting problem.

The first problem is solved by supersymmetry,
which can render the weak scale to be stable against
radiative corrections.
In SUSY GUTs the so called ``technical" solution \cite{tech}
of the second problem is based on `fine tuning':
after GUT gauge $G$ group breaking to
$G_{SM}=SU(3)_C\times SU(2)_W \times U(1)_Y$, by proper choice of the
tree level potential parameters, the mass of the doublet component may be
imposed to be zero (or of order $100$ GeV) while the mass of triplet
fragment is order $M_G$.
 Unfortunately this looks very unnatural.

The actual task is to obtain DT splitting without fine tuning.
Several possibilities were suggested in the literature:

a) In ``sliding singlet" model \cite{wit} the DT splitting occurs in exact
SUSY limit, but after SUSY breaking the hierarchy is spoiled \cite{pol}.

b) Missing partner mechanism \cite{dim} is based on purely group-theoretical
arguments and it can be implemented in SUSY $SU(5)$ theory.
By introducing the missing doublet multiplets:
$75+50+\overline{50}$ and
by employing several symmetries one can exclude the
direct mass term for the Higgs superfields $5+\bar 5$, containing
the Higgs doublets.
Consequently these doublets remain massless while their triplet partners
acquire large ($\sim M_G$) masses by the mixing with the massive
triplets from $50+ \overline{50}$. However, besides the unpleasant
fact that this mechanism employs huge representations, it can be
spoiled by the possible nonrenormalizable terms which are permitted
by all symmetries.

c) Missing VEV mechanism \cite {dim1}
closely resembles the missing partner mechanism and it can be realized
in SUSY $SO(10)$ within the basic multiplets: $45+10+10'$.
If~ $45$~ has VEV in B$-$L direction, then the $10\cdot 45\cdot 10'$ term
in the superpotential renders the Higgs doublets massless,
while the triplets acquire the large masses $\sim M_G$.
These solutions also suffer instability against the higher order terms.
It was shown \cite {babu}  that DW mechanism can be protected against the
effects of
higher-dimension operators, but very complicated field content is required.

d) The Goldstone boson mechanism \cite {ino,ber,ber1}
 in which the Higgs doublets are
identified with the pseudo-Goldstone bosons (PGB) of
the spontaneously broken global pseudosymmetry looks very promising.
The first attempts were done in SUSY $SU(5)$ with the
superpotential having the larger $SU(6)$ global symmetry \cite {ino}.
Implementing this idea in a more consistent way, the models were built
based on the $SU(6)$ gauge group \cite {ber,ber1}. In these models
Higgs doublets emerge as the PGB modes due to accidental
global $SU(6)\times SU(6)$ pseudosymmetry of the Higgs
superpotential.
It was shown in ref. \cite {ber1} that in fact the $SU(6)$ SUSY GUT
(or maybe its trivial unitary extensions to $SU(6+N)$)
are only viable possibilities for the Goldstone boson mechanism.
In the same paper by introducing additional discrete symmetries there
was constructed models in which higher order terms are harmless and
can not spoil the hierarchy.

e) The ``custodial" symmetry mechanism \cite {gia} also requires the $SU(6)$
gauge group. The Higgs doublet is
light since it is related by the custodial symmetry to another
doublet which after GUT symmetry breaking becomes an unphysical Higgs.
If direct mass term for the $35$-adjoint representation is omitted the
superpotential is flat and the mass term of the doublets is exactly
canceled in the exact SUSY limit. After SUSY breaking due to soft
terms cancelation is partial and doublets
acquire the masses of order $m_{3/2}$.
Because of this the intermediate scale is arising in the theory.
In general nonrenormalizable terms also may destroy the hierarchy.

In this paper we suggest the natural mechanism for the DT splitting
in SUSY $SU(6)$ theory.
This mechanism actually lies towards the lines
of ``custodial" symmetry mechanism \cite {gia},
but resembles the missing partner mechanism because of peculiarity
of the scalar content which we consider. For the $SU(6)$ symmetry
breaking the 175 representation is used which does not contain
the electroweak doublet fragments.
Two models are constructed in which
crucial role in DT splitting is played by discrete or
continuous $R$ symmetries in such a way as all dangerous higher order
operators are excluded from the superpotential.

\section{General idea: 175-plet and its properties}

$SU(6)$ is the minimal semi-simple group whose adjoint
representation contains  $G_{SM}$ Higgs doublets.
subgroup has the form $35=1+5+\bar 5+24$, where a
pair of doublet-antidoublet is contained in $5$ and $\bar {5}$
respectively. When the $SU(6)$ symmetry breaks down to $G_{SM}$
spontaneously, then one pair of doublets from the
scalar fields emerge as goldstone bosons eaten by the
corresponding gauge fields of $SU(6)$ which become massive through
the Higgs mechanism.  If there exists another pair
of the Higgs  doublets in the theory which after the $SU(6)$ breaking
remain massless
due to some mechanism \cite{ber,ber1,gia} while the color
triplets acquire masses of order $M_{G}$, then in the effective low
energy theory we will have one pair of light Higgs doublets as in the
minimal supersymmetric standard model.

In a number of $SU(6)$ models the
$35=\Sigma^{A}_{B}$ adjoint IRREP was used for the symmetry
breaking \cite{ber,ber1,gia}.
In exact SUSY limit the potential of 35-plet in general has three
degenerate minima corresponding to three following channels
of the $SU(6)$ symmetry breaking:
\begin{eqnarray}
a) & & SU(5)\times U(1) \nonumber \\
b) & & SU(4)\times SU(2)\times U(1) \\
c) & & SU(3)\times SU(3)\times U(1) \nonumber
\end{eqnarray}
In $(1,a)$ and $(1,b)$ cases the
$5+\bar 5$ and $(4,\bar 2)+(\bar 4,2)$ fragments from $\Sigma$ are
absorbed by the appropriate gauge fields.
Both this fragments contain the pair of $G_{SM}$
doublets.

Here we suggest a possibility to break the $SU(6)$ symmetry
by the selfconjugate IRREP --- 175: $175\equiv \Phi^{ABC}_{A'B'C'}$
 ($A,B,$... denote the $SU(6)$ indices)
which is full antisymmetric with respect to up and down
indices and $\sum \Phi^{ABC}_{AB'C'}=0$.

The decomposition of the 175-plet in terms of the $SU(5)$ subgroup
has the following form:
\begin{equation}
175=75+50+\overline {50}  \label {dec}
\end{equation}
Neither $75$ nor $50$ ($\overline {50}$) contains the Higgs doublet.
It is interesting to note that $75+50+\overline {50}$ are just these
IRREPs, which were needed for the ``missing partner" mechanism \cite {dim}
to be operative in SUSY $SU(5)$ model;
as we see by extending the $SU(5)$ group to $SU(6)$, it is
possible to put these IRREPs exactly in the one $SU(6)$ IRREP.

175 cannot break $SU(6)$ group into the
$(1,a)$ and $(1,b)$ channels, so that only the $(1,c)$ channel is possible.
With respect to the $SU(3)\times SU(3)\times U(1)$
subgroup $175$ decomposes into:
$$
175=(1, 1)_{0}+(1, 1)_{-6}+(1, 1)_{6}+(8, 8)_0+
(6, \bar 6)_2+(\bar 6,6)_{-2}+
$$
\begin{equation}
\\\ +(3, \bar 3)_2+(\bar 3, 3)_{-2}+(3, \bar 3)_{-4}+(\bar 3, 3)_4  \label
{175}
\end{equation}
where the subscripts denote the U(1) charges corresponding
to the $Y_{U(1)}=diag(1, 1, 1, -1,-1, -1)$ generator of $SU(6)$.
After the $175$ multiplet develops VEV in the
$SU(3)\times SU(3)\times U(1)$ singlet ---
$(1, 1)_0$ component, the $(3, \bar 3)_2+(\bar 3, 3)_{-2}$
fragments are absorbed by appropriate gauge bosons and become
unphysical.

Thus, the main feature of 175 IRREP is that it contains only
the $G_{I}=SU(3)_C\times SU(3)_W\times U(1)$ singlet
(among all maximal subgroups of $SU(6)$),
and thus instead of multidegenerate vacua we have only one
$SU(6)$-s breaking channel (1,c).
$SU(3)_C$
is ordinary color group and $SU(3)_W$ contains weak $SU(2)_W$ group.
If we introduce a pair of
sextet-antisextet scalar superfields $H$ and $\bar H$,
which develop VEVs on their sixth components,
than $SU(3)_W\times U(1)$ will break
to $SU(2)_W\times U(1)_Y$  group and doublet-antidoublet pair
from $H$ and $\bar H$, respectively will be absorbed. If in the
theory there exists another pair of sextet-antisextet superfields --
$H'+\bar H'$ and they are associated with $H+\bar H$ superfields by
global symmetry, e.g. $SU(2)_H$, then one pair of doublets
may remain massless after symmetry breaking.
As we will see later discrete or continuous $R$ symmetry is sufficient to
exclude
some unacceptable terms from the superpotential.

Thus, we choose the scalar content of our model as
$\Phi(175)+H_m(6)+\bar H^m(\bar 6)$,
where~ $m$~ is the~ $SU(2)_H$ index.
The 175 breaks $SU(6)$ down to the $G_I$
subgroup, and the $6+\bar 6$ fields break
$SU(3)_W\times U(1)$ to $SU(2)_W\times U(1)_Y$.
It is easy to check that
the VEV structure of $\Phi (175)$s singlet has the form:
\begin{eqnarray}
\Phi^{124}_{124}=\Phi^{125}_{125}=\Phi^{126}_{126}
=\Phi^{134}_{134}&=&\Phi^{135}_{135}=\Phi^{136}_{136}
=\Phi^{234}_{234}=\Phi^{235}_{235}=\Phi^{236}_{236}=-V  \nonumber\\
\Phi^{145}_{145}=\Phi^{146}_{146}=\Phi^{156}_{156}=
\Phi^{245}_{245}&=&\Phi^{246}_{246}=\Phi^{256}_{256}=
\Phi^{345}_{345}=\Phi^{346}_{346}=\Phi^{356}_{356}=~V   \nonumber\\
\Phi^{123}_{123}&=&-\Phi^{456}_{456}=3V                    \label {vac1}
\end{eqnarray}
where $1, 2, 3$ and $4, 5, 6$ respectively stand for the
$SU(3)_C$ and $SU(3)_W$ indices, while
for the VEVs of the sextet-antisextets we have:
\begin{equation}
<H_m>=<\bar H^m>=\left(
\begin{array}{cccccc}
0& 0& 0& 0& 0& v \\
0& 0& 0& 0& 0& 0 \end{array}         \label {vac2}
\right)
\end{equation}
$V$ and $v$ can be found from the potential.
{}From (\ref {vac1}) it is easy to
see, that VEV of any odd power of $\Phi $ is zero:
\begin{equation}
{\rm Tr<\Phi^{2N+1}>=0}    \label {tr}
\end{equation}
and consequently Tr$<\Phi^3>=0$. In order to obtain nonzero $V$ and $v$,
it is necessary to introduce also two gauge singlet superfields $s_1$ and
$s_2$.
Then the most general $SU(6)\times SU(2)_H$ invariant renormalizable
superpotential
\begin{equation}
W_1=(M_{\Phi}+\sigma_1 s_1+\sigma_2 s_2)\Phi^2+
(M_H+h_1 s_1+h_2 s_2)\bar H^m H_m +\lambda \Phi^3
+W_1(s_1, s_2)      \label {supgl}
\end{equation}
has the accidental global $SU(6)_\Phi \times U(6)_{H_1}\times U(6)_{H_2}
\equiv SU(6)_{\Phi }\times U(6)_H^2$ symmetry
under the independent rotation of $\Phi $ and $\bar H^m+H_m$ superfields.
The symmetry of the superpotential is higher then the symmetry of the
full Lagrangian. The VEV of $\Phi$ (eq. (\ref {vac1}))
breaks the $SU(6)_{\Phi}$ symmetry to
$SU(3)_{\Phi}\times SU(3)_{\Phi } \times U(1)_{\Phi}$.
Then $(3, \bar 3)_2+(\bar 3, 3)_{-2}$ fragments from $\Phi (175)$
become goldstone modes. The VEVs of $H+\bar H$  (eq. (\ref {vac2}))
break $U(6)_H^2$ to $U(5)_{H_1}\times U(6)_{H_2}$ and
two pairs of $5+\bar 5$
from $H_m+\bar H^m$ become massless Goldstones (The both pair remain massless
due to $SU(2)_H$ symmetry).
Thus due to the $SU(6)_\Phi \times U(6)_H^2$ global symmetry the
color triplets are left massless along with the doublets components.
In order to render the triplets massive, we have
to avoid this global symmetry. The one way to do so is to
include the higher order non-renormalizable terms in the superpotential.

In doing so, we observe that
although the $d=5$ term $\bar H^m \Phi^2 H_m$ violates
the $SU(6)_\Phi \times U(6)_H^2$ symmetry, it does not lead to the
desirable DT splitting. The reason is that the VEV of $\Phi^2$
can couple to $H,\bar H$ only in the $SU(6)$ singlet channel.
In expanded form this term reads:
\begin{equation}
 (\Phi^2)^{C}_{D}\bar H^{D,m}H_{C,m}=\Phi^{ABC}_{A'B'C'}\Phi^{A'B'C'}_
 {ABD}\bar H^{D,m}H_{C,m}     \label {0}
\end{equation}
while from (\ref {vac1}) we see that
$<\Phi^2>^C_D\sim V^2diag(1,~ 1,~ 1,~ 1,~ 1,~ 1)^C_D$.
Therefore, this structure gives the same contribution to the mass terms
of doublet and  triplet fragments in $H,\bar H$, and
if doublets remain massless, the mass of triplets also will vanish.
Thus we have to include in the superpotential the $d=6$ term
\begin{equation}
\bar H^m\Phi^3H_m          \label {1}
\end{equation}
One of the invariants (\ref {1}) has the form:
\begin{equation}
(\Phi^3)^{C}_{D}\bar H^{D,m}H_{C,m}=\Phi^{ABC}_{A'B'C'}
\Phi^{A'B'C'}_{A_{1}B_{1}C_{1}}\Phi^{A_{1}B_{1}C_{1}}_{ABD}\bar H^{D,m}H_{C,m}
\label {2}
\end{equation}
Using (\ref {vac1}) we see, that $<\Phi^3>^C_D\sim V^3diag(1, 1, 1, -1, -1,
-1)^C_D$.
Thus to obtain reasonable DT splitting it is necessary to include the (\ref
{1})
term  in the superpotential.

Among the quartic terms there is
$\alpha (\bar H^mH_n)(\bar H^nH_m)$ term, where brackets denote
summation by $SU(6)$ indices.
Expanding this term by $SU(2)_H$ indices we get:
\begin{equation}
(\bar H^mH_n)(\bar H^nH_m)=(\bar H^1H_1)^2+2(\bar H^1H_2)(\bar H^2H_1)+
(\bar H^2H_2)^2                       \label {3}
\end{equation}
If only first pair develops VEV, then the doublets
 which come from this pair are goldstone bosons.
First term from the (\ref {3}) gives contribution
to the mass of the first doublet-antidoublet pair, namely $2\alpha v^2$.
However the term (\ref {3}) does not take part in the formation of the mass
of the second doublet-antidoublet pair and consequently the mass
of the latter will be nonzero.
 If both pairs of scalar components from
sextet-antisextet superfields have nonzero VEVs,
there exists mixing between Higgs doublets.
One eigenvalue of the mass matrix corresponds
 to the eigenstate which is a goldstone boson.
Therefore nonlinear terms in $H$ (or $\bar H$) give
the different contributions to the
doublets' masses. In any case one doublet-antidoublet pair will be
goldstone, but the second pair of doublets will
have the undesirable mass (It is easy to verify, that another
quartic term $(\bar H^mH_m)^2$ does not spoil the hierarchy).
Therefore,
we must exclude the term (\ref {3}) by some symmetry reasons in such a way
as to keep the term (\ref {1}) and get the nonzero $V$ and $v$ in the limit
of unbroken SUSY.

Below we present two models which satisfy these conditions and
thus lead to the desirable DT splitting.

\section{Two Models}

{\bf Model 1.}
Let us introduce the discrete $R$-symmetry $Z_3$ under which all
scalar superfields and the superpotential
transform as
\begin{equation}
(\Phi, \bar H^m, H_m) \to e^{i\frac {2\pi}{3}} (\Phi, \bar H^m, H_m);
{}~~~~~~~ W\to e^{-i\frac {2\pi}{3}}W        \label {4}
\end{equation}
Then the most general $SU(6)\times SU(2)_H\times Z_3$ invariant
superpotential including the terms up to the fifth order has a form
\begin{equation}
W=M_H\bar H^mH_m+a\bar H^m\Phi^3H_m+M_\Phi\Phi^2+b\Phi^5      \label {5}
\end{equation}
where under $a$ and $b$ terms all possible contractions by $SU(6)$ indices
must be understood (we shall consider these terms in details later).

As it was mentioned above, the $\Phi (175)$ contains only $SU(3)_C\times
SU(3)_W\times U(1)$ singlet with (\ref {vac1})
vacuum structure. Thus it is impossible to change this direction
by (\ref {vac2}) structure because there does not exist another direction for
175 in the
group space. The $F_{H_{A,m}}$ term has the form:
\begin{equation}
F_{H_{A, m}}=M_H\bar H^{A, m}+a\bar H^{B,m}(\Phi^3)^A_B   \label {6}
\end{equation}
{}From $F_H=0$ condition with $<H>$ and $<\bar H>$ having the form (\ref
{vac2}),
we find
\begin{equation}
<\Phi^3>^6_6=-\frac {M_H}{a}               \label {7}
\end{equation}
and therefore
\begin{equation}
<\Phi^3>^A_B=\frac {M_H}{a}diag(1, 1, 1, -1, -1, -1)^A_B     \label {8}
\end{equation}

By substituting the $\Phi^3$ in (\ref {5})  by (\ref {8}), we can
calculate the masses of the
doublet and triplet components from the sextet-\-anti\-sextet
pairs. It is easy to
see that for doublets $M_D=0$ and for triplets
$M_T=2M_H$.

As we see two pairs of doublet-antidoublets remain massless, while the two
pairs of triplet-antitriplets have the masses of order $M_H$.
One pair of
doublet-antidoublet is absorbed by appropriate gauge fields and the second
one survives after the symmetry breaking.
There exists mixing between triplets (antitriplets) from $175$ and
triplets (antitriplets) from $H_1$ ($\bar H^1$). The mass matrix
for the triplet (antitriplet) components has
the following form:
{\vspace{5mm}}
\begin{eqnarray}
&\hspace{-6mm}3_{175}~~~~3_{H_1}~~~~3_{H_2}& \nonumber \\
\begin{array}{ccc}
\bar 3_{175} \\
\bar 3_{\bar H_1} \\
\bar 3_{\bar H_2}
\end{array}
&\hspace{-6mm}\left(
\begin{array}{ccc}
M_1& M_{12}& 0 \\
M_{12}& 2M_H& 0 \\
0& 0& 2M_H \end{array}
\right)&                                          \label {9}
\end{eqnarray}
where
\begin{equation}
det\left(
\begin{array}{cc}
M_1& M_{12} \\
M_{12}& 2M_H \end{array}
\right)=0                            \label {10}
\end{equation}
and thus one eigenvalue of the mass matrix is zero and the corresponding
eigenstate is a goldstone boson. Other two nonzero eigenvalues are $2M_H+M_1$
and $2M_H$; their magnitudes must
be not less than
$10^{16}~$ GeV
  to avoid fast proton decay due to the $d=5$ operators.
As we will see below, the $M_H$ can be compatible
with a proton stability. To demonstrate
this let us elaborate the superpotential in more details. After substituting
$\Phi $ and $H+\bar H$ in (\ref {5}) by (\ref {vac1}),
(\ref {vac2}) and using (\ref {tr}) all $b$ terms
become zero. There exist seven
possible $\bar H\Phi^3H$ invariants:
\begin{eqnarray}
I_1&=&\Phi^{ABC}_{A'B'C'}\Phi^{A'B'C'}_{A_{1}B_{1}C_{1}}\Phi^{A_{1}B_{1}C_{1}}_{ABD}\bar H^DH_C \nonumber \\
I_2&=&\Phi^{ABC}_{A_1B_1C_1}\Phi^{A_1A'B'}_{ABC'}\Phi^{B_{1}C_{1}D}_{CA'B'}\bar
H^{C'}H_D \nonumber  \\
I_3&=&\Phi^{ABC}_{A_1B_1C'}\Phi^{A_1A'B'}_{ABC_1}\Phi^{B_1C_1D}_{CA'B'}\bar
H^{C'}H_D \nonumber  \\
I_4&=&\Phi^{ABC}_{A'B'C'}\Phi^{A'B'C'}_{AA_{1}B_{1}}\Phi^{A_{1}B_{1}C_{1}}_{BCD}\bar H^DH_{C_1} \label{11} \\
I_5&=&\Phi^{ABC}_{A'B'C'}\Phi^{A'B'B_{1}}_{ABA_{1}}\Phi^{C'A_{1}D}_{CB_{1}C_{1}}\bar H^{C_{1}}H_{D} \nonumber\\
I_6&=&\Phi^{ABC}_{A'B'C'}\Phi^{A'B'C'}_{A_{1}B_{1}C_{1}}\Phi^{A_{1}B_{1}C_{1}}_{ABC}\bar H^DH_D \nonumber \\
I_7&=&\Phi^{ABC}_{A'B'C'}\Phi^{A'B_1C_1}_{ABA_1}\Phi^{B'C'A_1}_{CB_1C_1}\bar
H^DH_D  \nonumber
\end{eqnarray}
and thus
\begin{equation}
a\bar H^m\Phi^3H_m\equiv \sum^7_{i=1} a_{i}I_{i}       \label {12}
\end{equation}
where $a_i$ are the parameters of dimension GeV$^{-2}$:
$a_i=\frac {\lambda_i}{M^2}$ where $\lambda_i$ are
dimensionless parameters and $M$ is some cutoff mass parameter.
Using (\ref {11}), (\ref {12})
we get:
\begin{equation}
W(V,v)=M_Hv^2-48(36\lambda_1+2\lambda_2+\lambda_3+24\lambda_4+8\lambda_5)
\frac {V^3v^2}{M^2}+36^2M_{\Phi}V^2                          \label {13}
\end{equation}
The numbers are the combinator factors.
{}From the condition $F_V=F_v=0$ we have:
\begin{eqnarray}
V & = & M\left(
\frac{M_H}{M} \right)^{1/3} [48 (36\lambda_1+2\lambda_2+\lambda_3+
24\lambda_4+8\lambda_5)]^{-1/3} \nonumber  \\
v & = & 6M\left(
\frac {6M_{\Phi}^3}{M^2M_H}\right)^{1/6}
(36\lambda_1+2\lambda_2+\lambda_3+
24\lambda_4+8\lambda_5)^{-1/3}                              \label {14}
\end{eqnarray}
For $M\sim 10^{17}$~GeV (superstring scale), $M_H\sim 10^{16}~$GeV,
$M_{\Phi}\sim 10^{15}$~GeV,
$\lambda_i \sim 10^{-1}$ we get $V\sim v\sim 10^{16}$~GeV.
 Consequently, the respective {d=5} operator
is suppressed by the $M_H\sim 10^{16}$~GeV scale.

We have demonstrated how the cancelation
of doublets' masses occurs if the superpotential
has the form (\ref {5}). The higher
terms, which are permitted by the
$SU(6)\times SU(2)_H\times Z_3$ symmetry
and contain unacceptable
term (\ref {3}) are:
\begin{equation}
\frac {1}{M^5_{Pl}}\Phi^4\times (\bar H)^2\times (H)^2,~~~
\frac {1}{M^5_{Pl}}\Phi^2\times (\bar H)^3\times (H)^3,~~~
\frac {1}{M^5_{Pl}}(\bar H)^4\times (H)^4    \label {15}
\end{equation}
where all possible contractions by $SU(6)$ and $SU(2)_H$
indices is assumed. The doublet's
mass induced from these terms will have the magnitude
$M_D\sim M_{G}\left(\frac{M_{G}}{M_{Pl}}\right)^5$  and
for $M_{Pl}\sim 10^{19}$~GeV, we have $M_D\sim 10-100$ GeV,
which is indeed the desirable value for the ``$\mu $-term".

One may ask why the cut-off parameters $M$ and $M_{Pl}$
respectively in (\ref {13}) and (\ref {15}) have different magnitudes?
Let us assume that term (\ref {1})
is obtained by the heavy particle exchange mechanism \cite {fro}.
Let us introduce the pairs of vector supermultiplets: $\overline {210}^m+210_m$
($m$ is $SU(2)_H$ index). $210\equiv \Psi^{A'B'}_{ABC}$ is antisymmetric with
respect
to up and down indices and $\sum \Psi^{AB'}_{ABC}=0$.
If the superpotential for $\overline \Psi^m+\Psi_m$ has the form:
\begin{equation}
W_{\Psi}=M\overline {\Psi}^m \Psi_m+\alpha_1
\Phi \overline {\Psi}^mH_m+\alpha_2\Phi
\Psi_m\bar H^m+\alpha_3\Phi\overline {\Psi}^m\Psi_m    \label {Psi}
\end{equation}
the lowest operators which are obtained after integrating
 out the heavy $\overline \Psi^m+\Psi_m$ fields are
$\frac {1}{M}\bar H^m\Phi^2H_m$ and
$\frac {1}{M^2}\bar H^m\Phi^3H_m$.
It is easy to check that the higher operators
containing the combination (\ref {3}) will not be induced by
exchanges of $\Psi + \overline \Psi$.
The terms of eq. (\ref{15})
containing the multiplier (\ref {3}) may be induced only by the
nonperturbative gravity effects and thus they will be suppressed
by the Planck scale.

Note, that $W_{\Psi }$ do not has the definite $Z_3$ charge
as $W$  (see eq. (\ref {4})). To improve this drawback let us
introduce the gauge singlet superfield--- $s$ and instead of $Z_3$
symmetry $Z_8$, under which the scalar superfields and the
superpotential transform as:
$$
(\bar H^m, H_m, \overline \Psi^m, \Psi_m)\to e^{i\frac {2\pi }{8}}
(\bar H^m, H_m, \overline \Psi^m, \Psi_m),
{}~~~\Phi \to \Phi ,
$$
\begin {equation}
s\to e^{i\frac {2\pi}{4}}s;
{}~~~~~~~~~W\to e^{i\frac {2\pi }{4}}W  \label {16}
\end {equation}
were $W=W_1+W_{\Psi }$; $W_{\Psi}$ is given by (\ref {Psi}) and
\begin {equation}
W_1=M_H\bar H^mH_m+a\bar H^m\Phi^3H_m +\alpha s\Phi^2+
\frac {\beta}{M_{Pl}^2}s^5   \label {17}
\end {equation}
the lowest term, which contains multiplear (\ref {3}) and permitted by
these symmetries is:
$\frac {1}{M_{Pl}^4}\bar H^2\times H^2s^3$
from (\ref {17}) it is easy to verify that if
$\frac {\alpha}{\beta} \sim 10^{-4}$
then $<s>\sim 10^{16}$~GeV and we have $M_D\sim 10$~TeV.

{\vspace{3mm}}
{\bf Model 2.}
The second possibility to restrict all nonlinear
terms in $H$ (or $\bar H$), which in general may spoil the
hierarchy, is the continuous $R$-symmetry.
For example if $R_H=1$, $R_{\Phi }=R_{\bar H}=0$ (in the
units of the W-charge) then the superpotential will have the form:

\begin{equation}
W=\bar H^m(M_H+a_1\Phi^2+a_2\Phi^3+...)H_m   \label {19}
\end{equation}
Note that if $R_{\Psi }=R_H$ and $R_{\overline {\Psi }}=0$ then
the terms in $W_{\Psi}$
(see eq. (\ref {Psi})) have the same $R$ charge as $W$ (eq. (\ref {19})),
and thus they also can be incorporated in the theory.
{}From the condition $F_{H}=0$ the {VEV} of $\Phi$ is fixed. It
is possible to satisfy the $F_\Phi=0$
condition in such a way as to get nonzero VEV-s for $H (\bar H)$,
 if we arrange
their VEV-s as it was suggested in ref. \cite {gia}:
\begin{equation}
<H_1>=<\bar H^2>=(0, 0, 0, 0, 0, v); ~~~~~~~
<H_2>=<\bar H^1>=0     \label {20}
\end{equation}
It is obvious, that $F_\Phi=0$ condition is satisfied
and $v$ is undetermined. In other words,
(\ref {19}) has an $F$-flat and $D$-flat vacuum.

\section{Conclusions}

We have considered the SUSY $SU(6)$ theory in which the GUT symmetry
breaking occurs due to the Higgs superfield in the 175 representation.
It does not contain the Higgs doublet fragments and can break the
$SU(6)$ symmetry only in the
$SU(3)_C\times SU(3)_W\times U(1)$ channel.
By introducing also two pairs of the scalar superfields
in $6+\bar 6$ representation which are related
by the custodial global $SU(2)_H$ symmetry, we have constructed
two models in which the DT-splitting occurs naturally.
In these models
the discrete or continuous $R$ symmetries are used for
obtaining the desired structure of superpotential, which
along with the renormalizable terms
also include all allowed nonrenormalizable ones.
It is worth to stress that the latter also play a crucial
in providing the large masses to the Higgs triplets while for the
doublet components they may induce the desirable value for ``$\mu$-term".

The fermion sector of the model may be constructed in the same manner
as in ref. \cite {gia}, if
we have one $\bar 6^m+15$ ($m$ is $SU(2)_H$ index) anomaly-free
fermion supermultiplet per generation. The Yukawa superpotential which
generates up and down fermion masses has a form:
\begin{equation}
W_Y=g_d\bar 6^m~15~\bar H^n\epsilon_{mn} +
\frac {g_u}{M_0}15~15~H_m~H_n\epsilon_{mn}  \label {21}
\end{equation}
which implies the following assignment of charges:
$\bar 6^m\to e^{i\frac {2\pi}{3}}\bar 6^m$ in case of $Z_3$ symmetry
and $\bar 6^m\to e^{i\frac {2\pi}{8}}\bar 6^m$ in
case of $Z_8$ for the Model 1, and
$R_{15}=-R_H/2$ and $R_{\bar 6^n}=3R_H/2$ for the Model~2.
The last term in (\ref {21}) may be obtained by
exchange of the heavy superfields with mass of order $M_0$ \cite {fro}.
For the details concerning the fermion sector we refer to \cite {gia},
where this question is discussed.

Let us conclude with the following comment:
Neglecting the threshold effects we will have the
standard unification of the three gauge couplings.
Taking into account the threshold effects due to the
$75+50+\overline {50}$ multiplets, the picture will be changed.
It was shown in \cite {yam} that nonminimal (missing multiplet) $SU(5)$
model with these IRREPs allows small values of the
SUSY breaking scale $m_{susy}$ for any $\alpha_s (m_Z)$
in the experimentally allowed range. In the context
of our model this result cannot be directly applied
due to the different gauge sector.
The detailed study of the problem of the gauge coupling unification
in our model and its implications for the proton decay
is the subject of a separate investigation.

\vspace{5mm}

{\bf {Acknowledgements}}

We thank J.Chkareuli and G.Dvali for useful discussions. This work
was supported in part by the ISF grant $No.\rm MXL000$ and Georgian
Government and ISF grant $No.\rm MXL200$.

Z.T. thanks G.Senjanovic and especially Z.Berezhiani
 for useful discussions, also
the International Atomic Energy Agency and UNESCO for hospitality at
the International Center for Theoretical Physics, Trieste.

\end{document}